\documentclass[12pt]{article}

\makeatletter
\newcommand{\eqref}[1]{(\ref{#1})}
\newcommand{\Secref}[1]{\expandafter\MakeUppercase\secrefname~\ref{#1}}
\newcommand{\Appref}[1]{\expandafter\MakeUppercase\appendixname~\ref{#1}}
\newcommand{\Figref}[1]{\expandafter\MakeUppercase\figurename~\ref{#1}}
\newcommand{\Tbref}[1]{\expandafter\MakeUppercase\tablename~\ref{#1}}
   \let\secref\Secref  \let\appref\Appref
\let\figref\Figref  
\newcommand{\secrefname}{Section}
\DeclareSymbolFont{AMSb}{U}{msb}{m}{n}
\DeclareSymbolFontAlphabet{\mathbb}{AMSb}
\DeclareMathAlphabet\mathfrak{U}{euf}{m}{n}
\SetMathAlphabet\mathfrak{bold}{U}{euf}{b}{n}
\DeclareSymbolFont{boldletters}  {OML}{cmm} {b}{it}
\DeclareSymbolFont{boldoperators}{OT1}{cmr}{bx}{n}
\DeclareSymbolFont{boldsymbols}  {OMS}{cmsy}{b}{n}
\newcommand{\Vev}[1]{\Bigl\langle #1 \Bigr\rangle}
\newcommand{\vev}[1]{\langle #1 \rangle}
\newcommand{\inv}{^{\raise.15ex\hbox{${\scriptscriptstyle -}$}\kern-.05em 1}} 
\newcommand{\ext}[1][]{\mathop{\raisebox{.2ex}{$\textstyle\bigwedge$}^{#1}}}
\newcommand{\del}{\partial}
\newcommand{\e}{\mathrm{e}}                               
\newcommand{\dint}[2][]{\mathop{\mathalpha{\int#1}#2}}    
\newcommand{\doint}[2][]{\mathop{\mathalpha{\oint#1}#2}}  

\DeclareMathSymbol{*}{\mathop}{symbols}{"03} 
\newcommand{\ol}{\overline}
\newcommand{\ul}{\underline}
\newcommand{\group}[1]{\mathop{\kern\z@\mathrm{#1}}\nolimits}     

\newcommand{\lie}[1]{\mathop{\kern\z@\mathfrak{#1}}\nolimits}
\newcommand{\g}{\lie{g}}
\newcommand{\opname}[1]{\mathop{\kern\z@\mathrm{#1}}\nolimits}    

\newcommand{\eps}{\epsilon}
\newcommand{\CA}{\mathcal{A}}
\newcommand{\CB}{\mathcal{B}}
\newcommand{\CC}{\mathcal{C}}
\newcommand{\CD}{\mathcal{D}}
\newcommand{\CL}{\mathcal{L}}
\newcommand{\CM}{\mathcal{M}}
\newcommand{\CO}{\mathcal{O}}
\newcommand{\CS}{\mathcal{S}}
\newcommand{\CV}{\mathcal{V}}
\DeclareMathSymbol{\Ba}{\mathalpha}{boldletters}{`a}
\DeclareMathSymbol{\Bb}{\mathalpha}{boldletters}{`b}
\DeclareMathSymbol{\Bc}{\mathalpha}{boldletters}{`c}
\DeclareMathSymbol{\Bd}{\mathalpha}{boldletters}{`d}
\DeclareMathSymbol{\Bf}{\mathalpha}{boldletters}{`f}
\DeclareMathSymbol{\Bg}{\mathalpha}{boldletters}{`g}
\DeclareMathSymbol{\Bu}{\mathalpha}{boldletters}{`u}
\DeclareMathSymbol{\Bx}{\mathalpha}{boldletters}{`x}
\DeclareMathSymbol{\By}{\mathalpha}{boldletters}{`y}
\DeclareMathSymbol{\Bz}{\mathalpha}{boldletters}{`z}
\DeclareMathSymbol{\BF}{\mathalpha}{boldletters}{`F}
\DeclareMathSymbol{\BG}{\mathalpha}{boldletters}{`G}
\DeclareMathSymbol{\BQ}{\mathalpha}{boldletters}{`Q}
\DeclareMathSymbol{\BX}{\mathalpha}{boldletters}{`X}
\DeclareMathSymbol{\Bal}{\mathord}{boldletters}{"0B}
\DeclareMathSymbol{\Bbe}{\mathord}{boldletters}{"0C}
\DeclareMathSymbol{\Bga}{\mathord}{boldletters}{"0D}
\DeclareMathSymbol{\Bde}{\mathord}{boldletters}{"0E}
\DeclareMathSymbol{\Bxi}{\mathord}{boldletters}{"18}
\DeclareMathSymbol{\Bphi}{\mathord}{boldletters}{"1E}
\DeclareMathSymbol{\Bph}{\mathord}{boldletters}{"27}
\DeclareMathSymbol{\Bch}{\mathord}{boldletters}{"1F}
\DeclareMathSymbol{\Bps}{\mathord}{boldletters}{"20}
\DeclareMathSymbol{\Bom}{\mathord}{boldletters}{"21}
\DeclareMathSymbol{\BDe}{\mathalpha}{boldoperators}{"01}
\DeclareMathSymbol{\BPi}{\mathalpha}{boldoperators}{"05}
\DeclareMathSymbol{\BPs}{\mathalpha}{boldoperators}{"09}
\DeclareMathSymbol{\BV}{\mathord}{boldsymbols}{"34}
\DeclareMathDelimiter{\Blbracket}{\mathopen} {boldoperators}{"28}{largesymbols}{"00}
\DeclareMathDelimiter{\Brbracket}{\mathclose}{boldoperators}{"29}{largesymbols}{"01}
\newcommand{\Bracket}[1]{\Blbracket #1 \Brbracket}
\renewcommand{\section}{\@startsection{section}{1}{\z@}%
                                    {-7ex \@plus -1ex \@minus -.2ex}%
                                    {2.5ex \@plus.2ex}%
                                    {\normalfont\large\scshape\centering}}
\renewcommand{\subsection}{\@startsection{subsection}{2}{\z@}%
                                       {-5ex \@plus -1ex \@minus -.2ex}%
                                       {1.5ex \@plus.2ex}%
                                       {\normalfont\normalsize\scshape}}
\renewcommand{\subsubsection}{\@startsection{subsubsection}{3}{\z@}%
                                          {-4ex\@plus -1ex \@minus -.2ex}%
                                          {1.5ex \@plus .2ex}%
                                          {\normalfont\normalsize\scshape}}
\newcommand{\sectionname}{}

\renewcommand\@seccntformat[1]{\ignorespaces\csname #1name\endcsname\space
                               \csname the#1\endcsname.\quad}   
\renewcommand{\appendix}{\par
  \setcounter{section}{0}%
  \setcounter{subsection}{0}%
  \renewcommand{\thesection}{\@Alph\c@section}%
  \renewcommand{\sectionname}{\appendixname}}
\newdimen\captionmargin 
\newdimen\captionindent 
\newdimen\captionwidth 
\newcommand{\captionfont}{\slshape}
\newcommand\@captionlabel[1]{\textsc{#1:}\space}
\long\def\@makecaption#1#2{%
  \vskip\abovecaptionskip  
  \captionwidth\hsize 
  \advance\captionwidth -2\captionmargin
  \sbox\@tempboxa{\@captionlabel{#1}\captionfont #2}%
  \ifdim \wd\@tempboxa >\captionwidth
    \ifdim\captionindent>\z@ 
      \advance\captionwidth -\captionindent
      \hskip\captionindent
    \fi
    \hskip\captionmargin
    \parbox[t]{\captionwidth}{\leavevmode\hskip-\captionindent
      \@captionlabel{#1}\captionfont #2}%
  \else
    \global \@minipagefalse
    \hb@xt@\hsize{\hfil\box\@tempboxa\hfil}%
  \fi
  \vskip\belowcaptionskip}
\def\eqnarray{%
   \stepcounter{equation}%
   \def\@currentlabel{\p@equation\theequation}%
   \global\@eqnswtrue
   \m@th
   \global\@eqcnt\z@
   \tabskip\@centering
   \let\\\@eqncr
   $$\everycr{}\halign to\displaywidth\bgroup
       \hskip\@centering$\displaystyle\tabskip\z@skip{##}$\@eqnsel
      &\global\@eqcnt\@ne$\;\hfil{##}$\hfil
      &\global\@eqcnt\tw@$\;\displaystyle{##}$\hfil\tabskip\@centering
      &\global\@eqcnt\thr@@ \hb@xt@\z@\bgroup\hss##\egroup
         \tabskip\z@skip
      \cr
}
\ifcase \@ptsize
\or 
\or 
\fi
  \divide\@tempdima\baselineskip
  \@tempcnta=\@tempdima
\makeatother
\begin{document}

%
%
\thispagestyle{empty}

\begin{flushright}\scshape
YITP-SB-02-42, RUNHETC-2002-33\\
hep-th/0209214\\ 
September 2002
\end{flushright}
\vskip5mm

\begin{center}

{\LARGE\scshape
BV Quantization of Topological Open Membranes
\par}
\vskip15mm

\textsc{Christiaan Hofman$^{1,\dagger}$} and 
\textsc{Jae-Suk Park$^{2,3,\ddagger}$}
\par\bigskip
{\itshape
  ${}^1$New High Energy Theory Center, Rutgers University,\\
        Piscataway, NJ 08854, USA,\\
  \par\medskip
  ${}^2$Department of Physics, KAIST,\\
        Taejon, 305-701, Korea,
  \par\medskip
  ${}^3$C.N. Yang Inst. for Theoretical Physics \textnormal{and} Dept. of Mathematics,\\
  SUNY at Stony Brook, NY 11794, USA,}
\par\bigskip
\texttt{${}^\dagger$hofman@physics.rutgers.edu, ${}^\ddagger$jaesuk@muon.kaist.ac.kr}

\end{center}

\section*{Abstract}

We study bulk-boundary correlators in topological open membranes. 
The basic example is the open membrane with a WZ coupling to a 3-form. 
We view the bulk interaction as a deformation of the boundary string theory.  
This boundary string has the structure of a homotopy 
Lie algebra, which can be viewed as a closed string field theory. 
We calculate the leading order perturbative expansion of this structure. 
For the 3-form field we find that the C-field induces a trilinear 
bracket, deforming the Lie algebra structure. 
This paper is the first step towards a formal universal 
quantization of general quasi-Lie bialgebroids.

\newpage
\setcounter{page}{1}
%
%

\section{Introduction}
\label{sec:intro}

In this paper we study bulk-boundary correlators for topological 
open membrane models as discussed in \cite{js,tomalg}. 
These are basically deformed BF type theories on a 3-manifold with boundary, 
with a manifest BV structure implemented. The basic gauge fields, 
ghosts and antifields are combined into superfields, which can 
be understood as maps from the superworldvolume into a super target 
manifold $\CM$. These models will be referred henceforth as BV sigma 
models. Passing to a superworldvolume automatically describes 
differential form fields. The manifest BV structure will however make 
it rather straightforward to gauge fix the gauge theory. 

A particular example, and indeed the main motivation for this work, 
is the open 2-brane (according to the terminology of \cite{js}) 
with a topological WZ coupling to a closed 3-form. 
In \cite{js} the open 2-brane model was shown to 
correspond to a BV sigma model of BF type. This model can be 
viewed as the membrane analogue of the Poisson-sigma model 
\cite{schastro}. The latter model was studied by Cattaneo-Felder 
in \cite{cafe} to describe deformation quantization 
in terms of deformed boundary correlation functions of a 
topological open string theory. In fact it was shown in \cite{js}, 
that the open membrane model can be seen as a deformation of 
this model. We tackle the topological open 2-brane 
theory in this paper in a way similar to \cite{cafe}. 
The CF model captures the effect of a 2-form field background 
in string theory \cite{scho}, which gives rise to 
noncommutative geometry \cite{codo,seiwit}. 

This model could be a toy model for string theory in a background 
3-form field. Models for WZ couplings to a large 3-form field were 
studied in \cite{bebe,kasa,matshi,pio}, but rather from 
the point of view of the somewhat ill-defined boundary string theory. 
In \cite{bebe} it was shown that in a particular decoupling limit 
a stack of M5-branes in a 3-form field the open membrane action of 
the M2-brane reduces to such a particularly simple topological 
membrane model with a large C-field. Our model could perhaps 
shed some light on the role of the mysterious generalized 
theta parameter that is central in decoupling limits 
of open membranes \cite{bebe,om,bebe2,gentheta,schaar,bersch}. 
Admittedly, our treatment is perturbative in the 3-form, 
and therefore not able to directly describe this situation. 
In \cite{tomalg} we argue that in the context of our model, 
at least in some cases, a large 3-form can be related by a canonical 
transformation to a model with another value for the 3-form, which 
can be small. However this involves the choice of an auxiliary 
Poisson structure, whose interpretation is not clear to us at the moment. 

More general models in this class are defined in \cite{tomalg}. 
They are shown to describe deformations of so-called Courant 
algebroids \cite{cour,wein,royt} (which could also be called more descriptively 
quasi-Lie bialgebroids). These structures have a deep relation to 
problems of quantization. For example, the original exact 
Courant algebroid was developed as an attempt to geometrically 
describe quantization of phase space with constraints and gauge 
symmetries \cite{cour}. It is related to a mix of the tangent 
space and the cotangent space of a manifold. It can 
be deformed by a 3-form, which induces a deformation 
of the Poisson structure to a quasi-Poisson structure 
\cite{royt,sevwein,sev}. This induces a deformed deformation 
quantization. 

To explain some of the words above, 
let us start with a (quasi)-Lie bialgebra, also known equivalently 
as a Manin pair. It can be described in terms of a Lie algebra 
$\g$ and its dual space $\g^*$, such that the total space 
$\g\oplus\g^*$ is also a Lie algebra. Note that the latter 
has a natural inner product and $\g$ is a maximally isotropic 
Lie subalgebra. This is the formulation used for a Manin 
pair $(\g\oplus\g^*,\g)$. In the language of quasi-Lie bialgebras, 
the bracket on the total space is formulated in terms of 
extra structure on the Lie algebra $\g$: a 1-cocycle called the 
cocommutator $\delta:\g\to\ext[2]\g$ dual to the bracket 
restricted to $\g^*$, and an element $\varphi\in\ext[3]\g$, 
such that $\delta\varphi=0$ and $\delta^2=-[\varphi,\cdot]$. 
When $\g^*$ is also a Lie subalgebra or equivalently $\varphi=0$, 
$(\g\oplus\g^*,\g,\g^*)$ is called a Manin triple and $\g$ a 
Lie bialgebra. (Quasi-)Lie bialgebras are the infinitesimal 
objects corresponding to (quasi-)Hopf algebras \cite{drin1}. 
Next, an algebroid is a vector bundle $A$ with a Lie bracket 
on the space of sections, acting as differential operators of 
degree 1 in both arguments. A well known example of a Lie 
algebroid is the tangent bundle $TM$ of a manifold. When 
the base manifold is a point, the definition of a Lie algebroid 
simply reduces to that of a Lie algebra. A (quasi-)Lie bialgebroid 
combines the structure of (quasi-)Lie bialgebra and Lie algebroid. 
It is an algebroid $A$ which has a Lie bracket, a cocommutator 
and an element $\ext[3]A$ satisfying certain integrability relations. 
For a precise definition we refer to the literature, 
see \cite{royt} and references therein. 
There it was also shown that in general a quasi-Lie bialgebra 
is equivalent to the structure of Courant algebroid. Courant 
algebroids appeared originally as an attempt to geometrize 
general quantization of constraint gauge systems \cite{cour}. 
The particular model mentioned above, for the coupling to the 3-form 
field, is related to the exact Courant algebroid, for which $A=T^*M$. 

The correlation functions we will calculate can be understood 
as a deformation of a homotopy Lie ($L_\infty$) algebra, 
on the boundary of the membrane induced by the bulk couplings. 
The definition of this $L_\infty$ in terms of correlation functions 
was explained in \cite{homa2}. This $L_\infty$ structure can be 
identified with the structure of the closed string field theory 
of this boundary string. The $L_\infty$ structure of closed string 
field theory was demonstrated in \cite{zwie}. It was discussed 
in the context of topological strings in \cite{kvz,ksv}. 
It is indeed this $L_\infty$ structure that 
is naturally deformed by bulk membrane couplings \cite{homa2}. 
The semiclassical approximation of this $L_\infty$ structure 
is equivalent to the structure of quasi-Lie bialgebra and more generally 
a quasi-Lie bialgebroid, as explained in \cite{tomalg}. 
This structure is of course natural in string theory, 
and plays an important role in CS and WZW models. 

As the deformation to first order in the bulk couplings induce the 
quasi-Lie bialgebroid structure, we could expect that by taking higher 
order correlators into account we should find its quantization. 
For the ``rigid'' case of quasi-Lie bialgebras the quantization 
is a quasi-Hopf algebra \cite{drin1}. The boundary theory of CS theory, 
which is the WZW model, indeed has the structure of a quasi-Hopf algebra. 
In a subsequent paper \cite{tomquant} we will discuss an explicit 
construction of this quasi-Hopf algebra for our model. 
The existence of a universal quantization for Lie bialgebras 
was proven in \cite{etikaz}. More generally, this could be applied 
to the models related to genuine Courant algebroids. 
The path integral of the BV sigma models studied in this paper 
can be used to define a formal universal quantization, extended 
to quasi-Lie bialgebras and even quasi-Lie bialgebroids. 
The explicit quantization of the model discussed here will be 
an important first step in this quantization program. 

This paper is organized as follows. In the next section we 
review the basic structure of the BV sigma models for the 
topological open membranes. 
In \secref{sec:comp} we will perform the gauge fixing 
and calculate the propagators. In \secref{sec:interaction}
these will be used to calculate the bulk-boundary correlation functions 
relevant for the deformed $L_\infty$ structure of the boundary algebra. 
We conclude with some discussion.

\section{BV Action For the Open Membrane}
\label{sec:bvaction}

Here we shortly discuss topological open membrane models. 
We will only provide a sketch; more details can be found 
in \cite{tomalg,js}.

\subsection{BV Quantization}

We start by reviewing shortly the method of BV quantization 
\cite{batvilk1,batvilk2}. We will be very brief, and refer to the 
literature for more details, see e.g.\ \cite{henteit} for 
a good introduction. 

The models we will study in this paper will contain gauge fields. 
In order to properly define the path integral in this context 
we need to divide out the (infinite) volume of the gauge group, 
and we have to construct a well defined quotient measure 
for the path integral. The BV formalism is a convenient and 
general procedure to construct this measure. 

One first fermionizes the gauge symmetry, by introducing 
anticommuting ghost fields for all the infinitesimal 
generators of the gauge symmetry. The charge of the 
corresponding fermionic symmetry is the BRST charge $\BQ$. 
It squares to zero if the gauge symmetries close. 
There will be a corresponding charge $g$, called ghost number, 
such that the original fields have $g=0$ and the ghost 
fields $g=1$. Hence $\BQ$ has ghost number 1. The parity of 
the field will correspond to the parity of the ghost number. 
When there are relations between gauge symmetries one needs 
in addition also ghost-for-ghost fields with $g=2$, etcetera. 
All these fields will be referred to simply as ``fields''. 
In addition to these fields one needs to introduce corresponding 
``antifields''. They correspond to the equations of motion. 
Generally, if a field $\phi$ has ghost number $g$ then its antifield 
$\phi^+$ has ghost number $-1-g$. Also, in our conventions, the antifields 
of a $p$-form field will have form degree $d-p$ in $d$ dimensions. 
Regarding the ``fields'' and ``antifields'' 
as conjugate coordinates in an infinite dimensional phase space, 
one has a natural symplectic structure and a dual Poisson bracket 
$\Bracket{\cdot,\cdot}$. 
Due to the relation of the ghost numbers, the latter is an odd 
Poisson bracket and has ghost number 1. It is called the BV antibracket. 
Often we will just call it the BV bracket in this paper. 
Due to the odd degree, it is graded antisymmetric and it satisfies 
a graded Jacobi identity, 
\begin{eqnarray}
\Bracket{\alpha,\beta} &=& -(-1)^{(|\alpha|+1)(|\beta|+1)}\Bracket{\beta,\alpha},\\
\Bracket{\alpha,\Bracket{\beta,\gamma}} &=& \Bracket{\Bracket{\alpha,\beta},\gamma}
  +(-1)^{(|\alpha|+1)(|\beta|+1)} \Bracket{\beta,\Bracket{\alpha,\gamma} },
\end{eqnarray}
where $|\alpha|$ denotes the ghost number of $\alpha$. 
For BV quantization one requires in addition a BV operator $\BV$. 
This is an operator of ghost number 1 satisfying $\BV^2=0$ 
and is such that the BV bracket can be given by the failure for 
$\triangle$ for being a derivation of the product, 
\begin{equation}
  \Bracket{\alpha,\beta} = (-1)^{|\alpha|}\BV(\alpha\beta) 
-(-1)^{|\alpha|}\BV(\alpha)\beta  -\alpha\BV\beta. 
\end{equation}

At the linear level in antifields, the dependence of the action on the 
``antifields'' is determined by the gauge symmetry. More precisely, if 
$\phi_I^+$ is the antifield for the field $\phi^I$, the terms linear in the 
antifields are given by $S_1=\int\phi_I^+\BQ\phi^I$. Note that this 
implies that the gauge transformation of the fields can be recovered in terms 
of the corresponding Hamiltonian vector field, $\BQ\phi^I=\Bracket{S_1,\phi^I}$. 
More generally, the BV-BRST operator $\BQ$ is determined by the full 
BV action $S_{BV}$ by the relation $\BQ=\Bracket{S_{BV},\cdot}$. 
It squares to zero if the BV action satisfies the classical master equation 
$\Bracket{S_{BV},S_{BV}}=0$. Quantum mechanically this is modified 
to the quantum master equation $\Bracket{S_{BV},S_{BV}}-2i\hbar\BV S_{BV}=0$. 
This is equivalent to nilpotency of the quantum version of the 
differential, $\BQ-i\hbar\BV$. 
The Jacobi identity for the BV bracket implies that the BV-BRST 
operator is a graded derivation of the BV bracket. 

Let us now relate this to the path integral. A BV observable $\CO$ 
is a functional of the fields and antifields satisfying 
$\BQ\CO-i\hbar\BV\CO=0$. The expectation value of such an observable 
is calculated by the path integral 
\begin{equation}
  \vev{\CO} = \dint[_L]{\CD\phi} \e^{\frac{i}{\hbar}S_{BV}}\CO,
\end{equation}
where the integration is performed over a Lagrangian subspace 
$L$ in field space. The quantum master equation is equivalent to 
the condition that this expectation value does not change under 
continuous deformations of $L$ for any BV observable. A choice of 
Lagrangian subspace $L$ is called a gauge fixing. The Lagrangian 
subspace $L$ can be given in terms of a gauge fixing fermion $\Psi$, 
which is a function of the fields $\phi^I$ of ghost number $-1$. 
In terms of $\Psi$, the subspace $L$ is then given by fixing the 
antifields as $\phi_I^+=\frac{\del\Psi}{\del\phi^I}$. The quantum 
master equation then implies that the above expectation values 
are independent of continuous variations of $\Psi$. The idea is 
to choose $\Psi$ such that the kinetic terms in the action 
become nondegenerate, so that one can define a propagator 
and apply perturbation theory.

\subsection{Superfields and Action}

The topological open membranes we study are of BF types, that is, the fields 
are differential forms $\phi^I_{(p)}$ on the worldvolume $V$ of the membrane. 
An essential role will be played by the BRST operator $\BQ$ and 
the 1-form charge $\BG_\mu$, satisfying the crucial anti-commutation 
relations $\{\BQ,\BG_\mu\}=\del_\mu$. The existence of $\BG_\mu$ is guaranteed 
for any topological field theory, as the energy-momentum tensor is BRST exact. 
The above anti-commutation relation gives rise to descent equations 
for the observables. We will define descendants of operators by the recursive 
relation as $\CO^{(p+1)}=\BG\CO^{(p)}$. When the scalar operator $\CO^{(0)}$ 
is BRST closed, they satisfy the descent equation  $\BQ\CO^{(p+1)}=d\CO^{(p)}$. 

The theories we are interested in start from differential $p$-forms 
$\phi^I_{(p)}$ with gauge transformations giving a BRST operator of the form 
$\BQ\phi^I_{(p)}=d\phi^I_{(p-1)}+\cdots$, where the dots contain no derivatives. 
Possibly by introducing auxiliary fields, we can always choose the ghosts 
$\phi^I_{(p-1)}$ such that the field $\phi^I_{(p)}$ is its descendant. 
This can be extended to higher gauge symmetries. Higher descendants have 
negative ghost number, hence they will be antifields. We will consider 
all descendants found in this way as ``fundamental'' in our BV theory. 
It will be convenient to combine all these descendants---gauge fields, 
ghosts, and antifields--- into superfields, introducing anticommuting 
coordinates $\theta^\mu$ of ghost number $-1$,
\begin{equation}
  \Bphi^I(x,\theta) = \phi^I(x) + \theta^\mu \phi^{I(1)}_\mu(x)
 + \frac{1}{2} \theta^\mu\theta^\nu \phi^{I(2)}_{\mu\nu}(x)
 + \frac{1}{3!} \theta^\mu\theta^\nu\theta^\rho \phi^{I(3)}_{\mu\nu\rho}(x).
\end{equation}
On superfields we have $\BG_\mu=\frac{\del}{\del\theta^\mu}$. 
Viewing $(x^\mu|\theta^\mu)$ as coordinates on the supermanifold $\CV=\Pi TV$, 
where $\Pi$ denotes the shift of (ghost) degree in the fiber by $+1$, 
these superfields can be viewed as functions on this super worldvolume. 
They take value in some target superspace $\CM$. In this way superfields 
are maps between these supermanifolds, and we can formulate the model 
as a sigma-model with superspaces as target and base spacs. 
We will sometimes use the notation $\Bx$ for the collection of 
supercoordinates $(x^\mu|\theta^\mu)$.

For the general model, we start from a supermanifold $\CM$, with a symplectic 
structure $\omega$ of degree 2. This will be the target space of our sigma-model. 
Let $\phi^I$ denote a set of coordinates on $\CM$. They will induce 
superfields $\Bphi^I$ on the super worldvolume $\CV$. The set of 
supercoordinates form a map $\Bphi:\CV\to\CM$. The symplectic 
structure on $\CM$ induces a BV symplectic structure on superfield 
space given by 
\begin{equation}
  \Bom_{BV} = \int_\CV\Bphi^*\omega = \frac{1}{2}\int_\CV \omega_{IJ}\delta\Bphi^I\delta\Bphi^J,
\end{equation}
where the variations $\delta\Bphi^I$ can be understood as a basis of 
one-forms on superfield space. 
Note that this symplectic structure has ghost degree $-1$, due to the 
integration over the super worldvolume. 
This symplectic structure induces a twisted Poisson bracket of degree 1, 
or BV bracket, acting on functionals of the superfields. 
It is given by 
\begin{equation}
  \Bracket{\alpha,\beta} = \int_\CV\omega^{IJ}\frac{\del^R\alpha}{\del\Bphi^I}\frac{\del^L\beta}{\del\Bphi^J}
  \equiv \sum_{I,J,p}\int_V\omega^{IJ}\frac{\del^R\alpha}{\del\phi^I_{(p)}}\frac{\del^L\beta}{\del\phi^J_{(3-p)}},
\end{equation}
where the superscripts $R,L$ denote right and left derivatives, 
and $\omega^{IJ}$ is the inverse of $\omega_{IJ}$. 
It is easily seen that this bracket can be derived from a BV operator $\BV$. 

The BV action functional will be given by 
\begin{equation}\label{bvact}
  S_{BV} = \int_\CV \biggl( \frac{1}{2}\omega_{IJ}\Bphi^I \Bd\Bphi^J +\Bga \biggr),
\end{equation}
where $\Bga=\Bphi^*\gamma$ for $\gamma$ a function on $\CM$. The kinetic terms 
in this action have the usual BF structure $\int BdA$, with the ``$B$'' and ``$A$'' 
fields residing in conjugate superfields with respect to the BV structure. 
For $\gamma=0$ the BV-BRST operator is given by $\BQ_0=\Bd$. 
This satisfies the correct anticommutation relation with 
$\BG_\mu=\frac{\del}{\del\theta^\mu}$. In fact the form \eqref{bvact} of the action 
is essentially the only one consistent with this requirement. In order for 
it to satisfy the BV master equation, the function $\gamma$ should satisfy a 
corresponding identity, and the superfields should satisfy appropriate boundary conditions. 
These conditions can be described as follows. We denote by $[\cdot,\cdot]$ 
the Poisson bracket on the space $C^\infty(\CM)$ dual to the symplectic 
structure $\omega$. Similarly we can construct a BV-like second order differential 
operator $\triangle=\frac12\omega^{IJ}\frac{\del^2}{\del\phi^I\del\phi^J}$ for this bracket.  
Note that the BV bracket and BV operator on superfield space can be related
from these structures on $\CM$ by pullback. Then the BV master equation gives 
the condition $[\gamma,\gamma]+2i\hbar\triangle\gamma=0$. In addition there 
are conditions coming from the boundary terms. 
When restricted to the important set of operators of the form $\Bf=\Bphi^*f=f(\Bphi)$ 
for a function $f$ on $\CM$, we can express the deformed operator $\BQ$ in terms 
of the deformation $\gamma$ and the poisson bracket $[\cdot,\cdot]$ as 
$\BQ\Bf=\Bd\Bf+\Bphi^*[\gamma,f]$. In other words, the deformation of the BRST 
operator acts on these functions essentially through the Poisson differential 
$Q=[\gamma,\cdot]$. 

To get good boundary conditions, we choose a Lagrangian submanifold 
$\CL\subset\CM$, and restrict $\Bphi$ to take values in $\CL$ on the boundary 
$\del\CV$.\footnote{Here $\del\CV=\Pi T(\del V)$ is the boundary of the 
super worldvolume. Note that it involves fixing both the commuting and 
anticommuting normal coordinate $(x_\perp|\theta_\perp)$.} 
The Lagrangian condition will guarantee that the action satisfies 
the master equation for $\gamma=0$, and reduces the more general master 
equation to an algebraic equation (from the worldvolume point of view) 
for $\gamma$. We can also add a boundary term, 
\begin{equation}
  S_{bdy} = \int_{\del\CV}\Bbe,
\end{equation}
where $\Bbe=\Bphi^*\beta$ and $\beta$ is a function on $\CL$.

\subsection{Courant Algebroid and 3-Form Deformations}

We will only consider models with fields of non-negative ghost number. 
This also means that the ghost number should not exceed 2 (as otherwise 
its anti superfield will have negative ghost number). There will be 
superfields of ghost number 0, which will be denoted  $\BX^i$, 
and their antifields have ghost number 2, and are denoted $\BF_i$. 
Furthermore there are superfields of ghost number one. They come in 
conjugate pairs $\Bch_a$ and $\Bps^a$, containing each others antifields. 
The expansions of these superfields will read 
\begin{eqnarray*}
\BX^i &=& X^i + \theta\cdot P^{+i}
 + \frac{1}{2} \theta^2\cdot \eta^{+i}
 + \frac{1}{3!} \theta^3\cdot F^{+i}_,\\
\BF_i &=& F_i + \theta\cdot \eta_{i}
 + \frac{1}{2} \theta^2\cdot P_{i} 
 + \frac{1}{3!} \theta^3\cdot X^+_{i},\\
\Bch_a &=& \chi_a + \theta\cdot A_{a}
 + \frac{1}{2} \theta^2\cdot B^+_{a}
 + \frac{1}{3!} \theta^3\cdot \psi^+_{a},\\
\Bps^a &=& \psi^a + \theta\cdot B^a
 + \frac{1}{2} \theta^2\cdot A^{+a}
 + \frac{1}{3!} \theta^3\cdot \chi^{+a},
\end{eqnarray*}
where we suppressed the worldvolume indices and their contractions. 
The ghost degree zero scalars $X^i$ are coordinate fields on some bosonic 
target space $M$. The ghost degree one fields $\chi_a$ form coordinates 
on the fiber of some odd fiber bundle $\Pi A$ over $M$, while their 
antifields $\psi^a$ are 
coordinates on the fiber of the dual fiber bundle $\Pi A^*$. 
The total target superspace can be identified with the twisted cotangent 
bundle $\CM=T^*[2](\Pi A)$.\footnote{Here $[p]$ denotes the shift of 
the (fiber) degree by $p$, equivalent to $\Pi^p$.} Note that the fiber 
of this cotangent bundle contains the conjugate $F_i$ to the base coordinates 
$X^i$ and the conjugate $\psi^a$ to the fiber coordinates $\chi_a$. 
A special case arises when we take $A=T^*M$, which leads to a 
so-called exact Courant algebroid. 
It was shown in \cite{js} that this particular model gives rise to the 
topological membrane coupling to a 3-form WZ term. 

As the full target space is a cotangent bundle, it has a natural symplectic 
structure. On the space of all superfields, this induces the following odd 
symplectic structure, 
\begin{equation}
\Bom_{BV} = \int_\CV \Bigl( \delta\BF_i\delta\BX^i + \delta\Bps^a\delta\Bch_a \Bigr),
\end{equation}
giving a BV bracket of the form 
\begin{equation}
\Bracket{\cdot,\cdot} = \int_\CV \biggl(\frac{\partial}{\partial\BX^i}\wedge\frac{\partial}{\partial\BF_i} 
 + \frac{\partial}{\partial\Bch_a}\wedge\frac{\partial}{\partial\Bps^a}\biggr).
\end{equation}
It is related to the BV operator 
$\BV = \int_\CV\Bigl(\frac{\del^2}{\del\BX^i\del\BF_i} + \frac{\del^2}{\del\Bch_a\del\Bps^a}\Bigr)$. 

The BV action of the deformed theory is given by 
\begin{equation}
S_{BV} = \int_\CV \Bigl( \BF_i \Bd\BX^i + \Bps^a \Bd\Bch_a +\Bga \Bigr). 
\end{equation}

The interactions we will consider will be of the form 
\begin{equation}
  \Bga = \Ba^i_a\BF_i\Bps^a
  + \Bb^{ia}\BF_i\Bch_a
  + \frac{1}{3!}\Bc_{abc}\Bps^a\Bps^b\Bps^c 
  + \frac{1}{2}\Bc_{ab}^c\Bps^a\Bps^b\Bch_c
  + \frac{1}{2}\Bc_{a}^{bc}\Bps^a\Bch_b\Bch_c.
\end{equation}
Here the coefficients $\Ba,\Bb,\Bc$ can be any functions of the degree zero superfields 
$\BX^i$. The deformation $\gamma$ should of course satisfy the master equation 
$[\gamma,\gamma]+2i\hbar\triangle\gamma=0$. 

The canonical example is the exact Courant algebroid, based on the cotangent 
bundle $A=T^*M$. Here the indices $a$ and $i$ can be identified. 
We then take the coefficient of the $\psi F$ term to 
be $a^i_j=\delta^i_j$. This will generate the 
Lie bracket on vector fields. The main other interaction 
is the cubic interaction $\frac{1}{3!}c_{ijk}\psi^i\psi^j\psi^k$. 
Here $c$ should be a closed 3-form. More generally we can 
also turn on an antisymmetic bivector $b^{ij}$. 
The full deformation is then given by 
\begin{equation}\label{cgamma}
  \gamma = b^{ij}F_i\chi_j+\frac12(\del_kb^{ij}+b^{il}b^{jm}c_{klm})\psi^k\chi_i\chi_j
  + \frac12b^{il}c_{jkl}\psi^j\psi^k\chi_i+\frac16c_{ijk}\psi^i\psi^j\psi^k. 
\end{equation}
It satisfies the master equation provided that 
$c_{ijk}$ is a closed 3-form and 
$3b^{l[i}\del_l b^{jk]}+b^{il}b^{jm}b^{kn}c_{lmn}=0$. 
Notice that this is a deformation of the Poisson condition 
for the bivector $b^{ij}$. 

After integrating out the linearly appearing superfield $\BF$ 
it gives the Poisson sigma-model studied by Cattaneo-Felder \cite{cafe} 
on the boundary with a bulk membrane coupling to the 
3-form $c$ (by pull-back) \cite{js},
\begin{equation}
  \frac{1}{3!}\int_\CV \Bc_{ijk}\Bd\BX^i\Bd\BX^j\Bd\BX^k 
  + \int_{\del \CV}\Bigl(\Bch_i\Bd\BX^i +\frac{1}{2}\Bb^{ij}\Bch_i\Bch_j\Bigr).
\end{equation}
Hence this models is the basic example 
of a (topological) string deformed by the 3-form. 

Another special case arises when there are only multiplets of 
degree one. The only coefficients in the bulk terms are the 
$\Bc$'s above. The target space has the form $\CM=\Pi\g\oplus\Pi\g^*$, 
where $\g$ is the vector space associated to the $\Bch$ and 
$\g^*$ the dual vector space associated to $\Bps$. 
As shown in \cite{tomalg} the master equation is equivalent 
to the conditions that $\g$ is a quasi-Lie bialgebra---or 
equivalently it says that $(\g\oplus\g^*,\g)$ is a Manin pair. 
This is well known to be the infinitesimal structure of a 
quasi-Hopf algebra. We therefore expect to find this latter 
structure when quantizing the model.

\subsection{Boundary Conditions}

The boundary conditions are restricted by the following rules. First, the 
restriction to the boundary of $\BF_i\Bd\BX^i+\Bps^a\Bd\Bch_a$ should be zero. 
In general, the boundary condition for a field $\phi$ is the same as for the Hodge 
dual of its antifield, $*\phi^+$. Furthermore the boundary condition for 
a 1-form $\phi^{(1)}$ is the same as for $d\phi^{(0)}$, 
in order not to break the BRST invariance. 
What remains is to provide the boundary conditions of the scalars. We choose 
here for $X^i$ and $\chi_a$ Neumann boundary conditions, and for $F_i$ and $\psi^a$ 
Dirichlet boundary conditions. The boundary conditions are therefore 
\begin{equation}
\renewcommand{\arraystretch}{1.2}
\begin{array}{*3{r@{\;}c@{\;}l@{\qquad}}r@{\;}c@{\;}l}
d_\perp X &=& 0, & F &=& 0, & d_\perp\chi &=& 0, & \psi &=& 0, \\{}
P^+_\perp &=& 0, & \eta_{\parallel} &=& 0, & A_\perp &=& 0, & B_{\parallel} &=& 0, \\
(*\eta^+)_{\parallel} &=& 0, & (*P)_\perp &=& 0, & (*B^+)_{\parallel} &=& 0, & (*A^+)_\perp &=& 0.
\end{array}
\end{equation}
All the 3-forms are zero on the boundary. This will however not be relevant, 
as these will vanish after gauge fixing anyway. 
For the superfields $\BF$ and $\Bps$ these boundary conditions can be 
conveniently rephrased by saying that they vanish on the boundary $\del\CV$, 
i.e. at $(x_\perp|\theta_\perp)=0$. 
More generally, boundary conditions can be chosen by restricting the coordinate 
fields $\phi^I$ to map to a Lagrangian submanifold $\CL\subset \CM=T^*[2](\Pi A)$. 
Above we chose $\CL=\Pi A$. 

In the following we will call the fields $(X,\chi)$ living on the boundary 
the basic fields, and their anti-superfields $(F,\psi)$ conjugate fields.

\subsection{Boundary Observables, Boundary Algebra, and Correlators}

We can relate boundary observables to functions on the Lagrangian 
submanifold $\CL$  (equal to $\Pi A$ in the situation described in the last subsection). 
For $f\in \CB \equiv C^\infty(\CL)$ and a $p$-cycle $C_p\subset\del V$,
we can build the coordinate invariant integrated operators 
$\CO^{(p)}_{f,C_p}=\int_{C_p}f^{(p)}$. Let $x$ be a point on the boundary $\del V$. 
Then we have an operator from the scalar component of the superfield $\Bf=\Bphi^* f$, 
evaluated at $x=(x|0)$, 
\begin{equation}
  \CO_{f,x} = \phi^* f(x) = f(\phi(x)).
\end{equation}
If $C$ is a 1-cycle in $\del V$, let us denote by $\CC=\Pi TC$ the super 
extension in $\del\CV$. Then we have the first descendant operator  
\begin{equation}
  \CO^{(1)}_{f,C} = \oint_{\CC}\Bf = \oint_C f^{(1)}. 
\end{equation}
Note that the integral over $\CC$ includes an integration over 
$\theta$ in the tangent direction to $C$. 
Lastly, we have the operators for 2-cycles $S\subset\del V$, 
\begin{equation}
  \CO^{(2)}_{f,S} = \int_{\CS}\Bf = \int_S f^{(2)},
\end{equation}
where $\CS=\Pi TS$. For example, for the full boundary $S=\del V$, 
this corresponds to a deformation of the boundary interaction. 

The operators $\CO_{f,C_p}=\int_{\CC_p}\Bf$, with $\CC_p=\Pi TC_p$, 
are closed with respect to the undeformed BRST operator $\BQ_0=\Bd$, 
due to Stokes' theorem and the fact that $C_p$ has no boundary. 
Also note that the BV operator is manifestly zero on the boundary, 
due to the Lagrangian condition on $\CL$. 
More generally, the deformed BRST operator will act on $\Bf$ for $f\in\CB$ 
through the differential $Q=[\gamma,\cdot]$ restricted to the 
``boundary algebra'' $\CB$. Therefore the operators above are genuine 
observables as long as this differential vanishes. If $Q$ 
is nonzero on $\CB$, we will still loosely speak of the above operators as 
``observables'', even though they are not necessarily closed. Genuine observables 
should then be constructed from the $Q$-cohomology of $\CB$. 
A more extended discussion of observables, especially relevant for nonzero $Q$, 
is beyond the scope of this paper and will appear elsewhere \cite{tomquant}. 

We will be interested in the effect of the bulk terms on the string 
theory living on the boundary. This topological closed string field 
theory has the structure of a $L_\infty$ algebra \cite{zwie,ksv,kvz}, 
generating its closed string field  theory. The bracket in 
this $L_\infty$ algebra is defined as the current algebra bracket 
of the boundary string, 
\begin{equation}\label{bracket}
  \{f,g\}=\frac{1}{i\hbar}\doint[_C]{f^{(1)}} g, 
\end{equation}
where $C$ is a 1-cycle on the boundary enclosing the insertion 
point of $g$ on the boundary. This bracket can more concretely be 
calculated using the correlation functions. More generally, 
the $L_\infty$ brackets can be defined by the correlation functions \cite{homa2} 
\begin{equation}\label{brackets}
  \Vev{\CO_{\delta_{\phi_0},\infty}\CO_{\{f_1,\ldots,f_n\},x}} = 
  \frac{(-1)^{\sum_k (n-k)(|f_k|+1)}}{(i\hbar)^{n-1}}\Vev{\CO_{\delta_{\phi_0},\infty}\CO^{(1)}_{f_1,C}\CO_{f_2,x}\CO^{(2)}_{f_3,\del V}\cdots\CO^{(2)}_{f_n,\del V}},
\end{equation}
where $C$ is a 1-cycle enclosing the point $x$. The powers of $i\hbar$ are 
included for convenience to cancel the leading behavior. 
The first insertion is a delta-function $\delta_{\phi_0}(\phi)=\delta(\phi-\phi_0)$ 
inserted in a point at ``infinity''. This outgoing test-observableis inserted 
to give an expectation value $\phi_0$ to the scalar fields living on 
the boundary. In most of the rest of this paper we will have the insertion 
of this operator understood, and will not write it down explicitly.

Let us first discuss the correlation functions of boundary operators in the 
open membrane theory, in the presence of a nontrivial bulk term $\gamma$. 
As we discussed above, the basic boundary observables are determined by 
functions on the Lagrangian subspace $\CL\subset\CM$. The bulk observables 
are induced by elements of the bulk algebra $\CA=C^\infty(\CM)$, while the 
boundary observables are induced by elements of the boundary algebra 
$\CB=C^\infty(\CL)$. Furthermore we have the projection $P_\CL:\CA\to\CB$, 
restricting a function to $\CL$. 

First we write the action as the sum of a kinetic term and an interaction term, 
$S=S_0+S_\gamma$, where we took $S_\gamma=\int\Bga$. 
This gives rise to the path integral representation of the correlation functions 
\begin{equation}
  \Vev{\prod_a\CO_a} = 
 \dint{\CD\Bphi}\e^{\frac{i}{\hbar}(S_0+S_\gamma)}\prod_a\CO_a,
\end{equation}
which we calculate as usual in an $\hbar$ expansion by perturbation theory, 
treating $S_\gamma$ as a perturbation. 
The propagator has the form 
\begin{equation}
\vev{\Bphi^I(\Bx)\Bphi^J(\By)} = -i\hbar\omega^{IJ}\mathbf{G}(\Bx,\By),
\end{equation}
where $\mathbf{G}(\Bx,\By)$ is the integral kernel for the inverse kinetic operator $\Bd\inv$ 
(after gauge fixing) and $\omega^{IJ}$ is the inverse of the symplectic structure $\omega_{IJ}$. 
We recognize in this the BV bracket structure. Because of this we will see 
that we can effectively describe the algebraic structure on the boundary operators 
in terms of the original BV bracket. 

Let us consider for concreteness the bracket defined by the correlation 
function\footnote{Here and for the rest of the paper we will use a normalization 
such that $\vev{\CO_{f,x}}=\int f$.} 
\begin{equation}
  \{f,g\}(\phi_0) = \Vev{\CO_{\delta_{\phi_0},\infty}\CO^{(1)}_{f,C}\CO_{g,x}},
\end{equation}
where all the operators are put on the boundary and $\delta_{\phi_0}$ is a delta function 
fixing the scalar fields to a fixed value $\phi_0$ consistent with the boundary condition. 
After contractions, and using the expression for the propagator above, the lowest 
order term can be written, 
\begin{equation}
  \pm\dint[_\CV]{d\Bz}\doint[_\CC]{d\By}\mathbf{G}(\Bz,\By)\mathbf{G}(\Bz,\Bx)
  \dint[_\CM]{d\phi}\delta(\phi-\phi_0)\omega^{KL}\omega^{IJ}\frac{\del^2\gamma}{\del\phi^K\del\phi^I}\frac{\del f}{\del\phi^J}\frac{\del g}{\del\phi^L}. 
\end{equation}
This is just the Feynman integral corresponding to a 2-legged tree-level diagram. 
The integral is a universal factor, which no longer depends on the precise choice 
of operators. The dependence on the functions $f$ and $g$, and therefore the choice 
of boundary observables, is expressed in terms of differential operators acting on 
these functions. 

In terms of the the boundary algebra of functions $\CB=C^\infty(\CL)$, 
the bracket can now be written 
\begin{equation}
  \{f,g\} = (-1)^{|f|+1}P_\CL[[\gamma,f],g] -(-1)^{(|f|+1)|g|}  P_\CL[[\gamma,g],f].
\end{equation}
Here the $P_\CL$ results from the integration against the outgoing state $\delta_{\phi_0}$, 
or equivalently the delta-function in the integral over zero-mode $\phi$. 
More precisely, we should interpret the boundary operators like $f$ as 
embedded in the algebra $\CA$; so we should more properly use a lift 
$f$ to the bulk algebra. 

Similarly, the 4-point function, defined by 
\begin{equation}
  \Vev{\CO_{\delta_{\phi_0},\infty}\CO^{(2)}_{h,\del V}\CO^{(1)}_{g,C}\CO_{f,x}},
\end{equation}
at tree level is proportional to the Feynman integral 
\begin{equation}\label{bndcorr}
  \dint[_\CV]{d\Bu}\dint[_{\del\CV}]{d\Bz}\doint[_\CC]{d\By}\mathbf{G}(\Bu,\Bz)\mathbf{G}(\Bu,\By)\mathbf{G}(\Bu,\Bx), 
\end{equation}
multiplied by a 3-differential operator acting on $f,g,h$ 
and depending on $\gamma$. The Feynman integral calculates 
again the universal coefficient corresponding to this term in 
the expansion of the trilinear bracket. The rest can again 
be expressed in terms of $\gamma$ and the BV bracket $[\cdot,\cdot]$, as 
\begin{equation}
  \{f,g,h\} = (-1)^{|g|+1}P_\CL[[[\gamma,f],g],h] \pm perms.
\end{equation}
The signs are such that the bracket is skew symmetric with respect 
to the ghost degree shifted by one. 

We see in general that the integrals over propagators give some universal 
coefficients, while the rest is determined by the algebra of the bracket. 
The essential point is that the nontrivial operations, i.e. the brackets 
defined above, correspond to nonvanishing Feynman integrals.

\section{Sigma Model Computations}
\label{sec:comp}

In this section we will compute the propagators using the BV quantization of the sigma-model. 

\subsection{Gauge Fixing}

The BV model having form fields, will have gauge invariance. We therefore 
need to gauge fix. The BV language we have adopted will make this quite simple. 
We mainly have to choose a gauge fixing fermion $\Psi$ to gauge fix the anti-fields. 
Note that in order to preserve the topological nature of our model we need 
to choose the fields and anti-fields according to ghost number: the anti-fields 
are the fields with negative ghost number. 

There were two types of ``BV multiplets'': $X^i$ and $F_i$ having degree 0 and 2, 
and $\chi_a$ and $\psi^a$, both having degree 1. The fields will have 
different degrees in the two cases, so the gauge fixing will be slightly different. 
We will therefore treat them separately. We will leave out the indices, as they 
can be easily reinserted. 

\subsubsection{Ghost Degree 1 Multiplet}

We start with the ghost degree 1 multiplets $(\Bch,\Bps)$. The gauge fields are 
1-form fields $A$ and $B$. We will use a covariant Lorentz gauge. 
To implement this gauge fixing, we introduce antighost fields 
and Lagrange multiplier fields. They both are scalars and have ghost numbers 
$-1$ and 0 respectively. 
They are of course supplemented by their antifields, 
which are 3-forms of ghost number 0 and $-1$ respectively. 
For the gauge field $A$ we have an antighost $\ol\chi$ and Lagrange multiplier 
$\ul\chi$; for $B$ the antighost is $\ol\psi$ and the Lagrange multiplier 
$\ul\psi$. The boudnary conditions for the antighosts and Lagrange multipliers 
will be the same as for the scalar field in the corresponding superfield. 

To fix the gauge we introduce the following antighost terms in the action, 
\begin{equation}
  S_{antighost} = \int\Bigl(\ul\chi\ol\chi^+ + \ul\psi\ol\psi^{+} \Bigr).
\end{equation}
The gauge fixing fermion will be given by 
\begin{equation}
  \Psi = \int \Bigl(d\ol\chi*A + d\ol\psi*B\Bigr).
\end{equation}
This implies the following gauge fixing of the antifields 
\begin{equation}
 A^{+} = *d\ol\chi,\qquad B^+ = *d\ol\psi,\qquad \ol\chi^+=-d*A,\qquad \ol\psi^+=-d*B, 
\end{equation}
while all other antifields vanish. 
The antifields of the Lagrange multipliers all vanish. 
After gauge fixing, the kinetic terms in the action $S_0$ become 
\begin{equation}
S_{kin} = \int_V \Bigl(
 BdA + A*d\ul\chi + B*d\ul\psi 
 + \chi d*d\ol\chi + \psi d*d\ol\psi \Bigr).
\end{equation}
These kinetic terms can be grouped into basically two multiplets: 
there are second order terms involving 2 scalars $(\chi,\ol\chi)$ and 
$(\psi,\ol\psi)$ and a set of 2 vectors and two scalars, $(A,B,\ul\chi,\ul\psi)$. 

In the following we will denote by $d_p:\Omega^p\to\Omega^{p+1}$ the 
De Rham differentials acting on $p$-forms  
and by $\Delta_p=-d_{p-1}d_p^\dagger-d_{p+1}^\dagger d_p$ the corresponding 
Laplacians acting on $\Omega^p$. 
The kinetic operator for two scalars is given by $\Delta_0=*d_2*d_0:\Omega^0\to\Omega^0$. 
This operator has finite dimensional kernel, and therefore can be inverted on the 
fluctuations. The kinetic operator for the vector `multiplet' can be conveniently 
organized in a matrix form as 
\begin{equation}
\pmatrix{*d_1 & d_0 \cr -*d_2* & 0}: \quad \Omega^1\oplus\Omega^0\to\Omega^1\oplus\Omega^0. 
\end{equation}
This matrix operator is a Dirac operator, in the sense that it squares to 
(minus) the Laplacian. This allows us to write the propagators in matrix 
notation as 
\begin{equation}
  \pmatrix{\vev{AB}& \vev{A\ul\chi}\cr \vev{\ul\psi B}& \vev{\ul\psi\ul\chi}}
  =i\hbar\pmatrix{*d_1 & d_0 \cr -*d_2* & 0}\inv =
  i\hbar\pmatrix{-*d_1\Delta\inv_{1,D}& -d_0\Delta\inv_{0,N}\cr *d_2*\Delta\inv_{1,D}& 0}. 
\end{equation}
The extra subscript on the inverse Laplacians denotes the boundary condition.

\subsubsection{Ghost Degree 0 Multiplet}

Next consider the $(\BX,\BF)$ multiplet. This one is slightly more 
complicated, as it involves a 2-form $P$ in $\BF$. Therefore we have to 
worry about an extra gauge-for-gauge symmetry. The antighost and Lagrange 
multiplier fields will be given by 
\begin{equation}
\renewcommand{\arraystretch}{1.2}
\begin{array}{c|ccc|ccc}
 \mbox{gauge~field} & \mbox{antighost} & \mbox{degree} & \mbox{ghost~\#} & \mbox{Lagr.~mult.} & \mbox{degree} & \mbox{ghost~\#} \\
\hline 
P    & \ol\eta & 1 & -1 & \ul\eta & 1 & 0  \\
\eta & \ol F   & 0 & -2 & \ul F   & 0 & -1 \\
\ol\eta & \ol\lambda & 0 & 0 & \ul\lambda & 0 & 1 
\end{array}
\end{equation}

The antighost terms in the action will be 
\begin{equation}
  S_{antighost} = \int\Bigl(\ul\eta\ol\eta^+ - \ul F\ol F^+ - \ul\lambda\ol\lambda^+\Bigr).
\end{equation}
and the gauge fermion is 
\begin{equation}
  \Psi = \int \Bigl(d\ol\eta*P + d\ol F*\eta + d\ol\lambda*\ol\eta\Bigr).
\end{equation}
The antifields will be replaced by the gauge fixing according to 
\begin{equation}
  P^+=*d\ol\eta,\qquad \eta^+=*d\ol F,\qquad \ol\eta^+=d*P+*d\ol\lambda,\qquad
  \ol F^+=-d*\eta,\qquad \ol\lambda^+=*d\ol\eta,
\end{equation}
and fixes all other antifields to zero. 
This gives the gauge fixed kinetic action 
\begin{equation}
  S_{kin} = \int_V \Bigl(
   Fd*d\ol F + PdX + P*d\ul\eta + \ul\eta*d\ol\lambda - \ol\eta*d\ul\lambda
   + \eta d*d\ol\eta + \eta*d\ul F  \Bigr).
\end{equation}
They split into a scalar multiplet $(F,\ul F)$, a 1-form multiplet 
$(*P,\ul\eta,X,\ol\lambda)$ (note that we dualized the 2-form $P$), and another
 1-form multiplet $(\eta,\ol\eta,\ul\lambda,\ul F)$ which has different kinetic terms. 
The scalar multiplet and the first 1-form  multiplet are handled in the 
same way as above, so we find the propagators 
\begin{equation}
  \pmatrix{\vev{*P\ul\eta}& \vev{*PX}\cr \vev{\ol\lambda\ul\eta}& \vev{\ol\lambda X}}
  = i\hbar\pmatrix{-*d_1\Delta\inv_{1,D}& -d_0\Delta\inv_{0,N}\cr *d_2*\Delta\inv_{1,D}& 0}.
\end{equation}

The fermionic 1-form multiplet needs extra attention. 
The kinetic operator of this multiplet is 
\begin{equation}
\pmatrix{-*d_1*d_1 & -d_0 \cr *d_2* & 0}:\quad \Omega^1\oplus\Omega^0\to\Omega^1\oplus\Omega^0. 
\end{equation}
Similar to the above we find the propagator 
\begin{equation}
  \pmatrix{\vev{\eta\ol\eta}& \vev{\eta\ul F}\cr \vev{\ul\lambda\ol\eta}& \vev{\ul\lambda\ul F}}
  = i\hbar\pmatrix{-*d_1*d_1 & -d_0 \cr *d_2* & 0}\inv
  = i\hbar\pmatrix{-*d_1*d_1\Delta_{1,D}^{-2}& d_0\Delta\inv_{0,D}\cr -*d_2*\Delta\inv_{1,D}& 0}. 
\end{equation}

\subsection{Explicit Propagators and Superpropagators}

To give explicit expressions for the propagators, we take for the membrane simply the 
upper half space. We choose coordinates $(x^\alpha,x_\perp)$, $\alpha=1,2$, with the boundary 
at $x_\perp=0$, and the bulk at $x_\perp>0$. We define reflected coordinates $\tilde x^\mu$ 
such that $\tilde x^\alpha=x^\alpha$, $\tilde x_\perp=-x_\perp$. We will also introduce a reflected 
Kronecker $\delta$ such that $\tilde\delta^{\perp\perp}=-1$, $\tilde\delta^{\alpha\alpha}=1$. 

We will denote the kernels of the inverse Laplacians $\Delta\inv_{p,B}$ by $\Pi^{p,B}$. 
Here $p$ is the form degree, and $B\in\{D,N\}$ denotes the boundary condition. 
The propagator of the scalar $\chi$ and $\ol\chi$, which have Neumann boundary conditions, 
is given by 
\begin{equation}
  \vev{\chi(x)\ol\chi(y)} = i\hbar\Pi^{0,N}(x,y) = 
  -\frac{i\hbar}{4\pi}\biggl(\frac{1}{\|x-y\|} + \frac{1}{\|x-\tilde y\|}\biggr).
\end{equation}
Similarly, there is a minus sign in between the two terms for Dirichlet boundary conditions,
\begin{equation}
  \vev{\psi(x)\ol\psi(y)}  = \vev{F(x)\ol F(y)} = i\hbar \Pi^{0,D}(x,y) 
  = -\frac{i\hbar}{4\pi}\biggl(\frac{1}{\|x-y\|} - \frac{1}{\|x-\tilde y\|}\biggr).
\end{equation}

The kernel for the inverse Laplacian $\Delta_{1,N}\inv$ for two vectors 
with Neumann boundary conditions is given by 
\begin{equation}
  \Pi^{1,N}_{\mu\nu}(x,y) = -\frac{1}{4\pi}\biggl(\frac{\delta_{\mu\nu}}{\|x-y\|} + \frac{\tilde \delta_{\mu\nu}}{\|x-\tilde y\|}\biggr),
\end{equation}
while $\Pi^{1,D}(x,y)$ has a minus sign in front of the reflected term. 
The propagator between the two vectors $A$ and $B$ can then be written 
\begin{equation}
  \vev{B_\mu(x)A_\nu(y)} 
  = i\hbar\eps_{\mu\rho\sigma}\frac{\partial}{\partial x^\sigma}\Pi^{1,N}_{\rho\nu}(x,y)
  = i\hbar\eps_{\nu\rho\sigma}\frac{\partial}{\partial y^\sigma}\Pi^{1,D}_{\rho\mu}(x,y).
\end{equation}
Notice that this indeed satisfies the boundary conditions for $A$ and $B$. 

The propagator between a vector and a Lagrange multiplier scalar, having always the 
same boundary condition, is given by the formula $d_0\Delta_0\inv$, where the 
propagator $\Delta_0\inv$ of course is chosen for the correct boundary conditions.
For example, 
\begin{equation}
  \vev{*P_\mu(x)X(y)} = -i\hbar d_0\Pi^{0,N}(x,y) 
  = -\frac{i\hbar}{4\pi}\biggl(\frac{(x-y)^\mu}{\|x-y\|^3} + \frac{(x-\tilde y)^\mu}{\|x-\tilde y\|^3}\biggr), 
\end{equation} 
indeed satisfying Neumann boundary conditions.

In practice we will not need all the propagators. For the calculations in this paper
we can ignore the Lagrange multiplier fields. In fact, all we need are the 
components of the gauge fixed superfields, which are 
\begin{equation}
\renewcommand{\arraystretch}{1.2}
\begin{array}{r@{\;}c@{\;}l@{\qquad}r@{\;}c@{\;}l}
\BX(x,\theta)  &=& X+\theta\cdot*d\ol\eta+\frac12\theta^2\cdot*d\ol F, &
\BF(x,\theta)  &=& F+\theta\cdot\eta+\frac12\theta^2\cdot P, \\
\Bch(x,\theta) &=& \chi+\theta\cdot A+\frac12\theta^2\cdot*d\ol\psi, &
\Bps(x,\theta) &=& \psi+\theta\cdot B+\frac12\theta^2\cdot*d\ol\chi,
\end{array}
\end{equation}

One can combine the above propagators in terms of a superpropagator. 
Let us define the superpropagator more generally. We introduce the 
supercoordinates $\Bx=(x^\mu|\theta^\mu)$ and $\By=(y^\mu|\zeta^\mu)$. 
Next we combine the propagators of the $p$-forms into a single superpropagator, 
\begin{equation}
  \sum_{p=0}^2\frac{1}{p!(2-p)!}\theta^{2-p}\cdot*d_p\Pi^p(x,y)\cdot\zeta^p.
\end{equation}
Here $\Pi^p=*\Pi^{3-p}*$. We express this in terms of a superpropagator 
\begin{equation}
  \BPi(\Bx,\By) = \sum_{p=0}^3\frac{1}{p!(3-p)!}\theta^{3-p}\cdot*\Pi^p(x,y)\cdot\zeta^p,
\end{equation}
and the operator $\Bd^\dagger$ representing  $d^\dagger$, that is 
$\frac{1}{p!}\Bd^\dagger(\theta^p\cdot\alpha_p) = \frac{1}{(p-1)!}\theta^{p-1}\cdot d^\dagger_p(\alpha_p)$. 
We then find that the above superpropagator can be written in the form 
$\Bd^\dagger\BPi(\Bx,\By)$. Note that the superpropagator $\BPi$ 
can be seen as the inverse of the super-Laplacian 
$\BDe=-\Bd\Bd^\dagger-\Bd^\dagger \Bd$, as it satisfies 
\begin{equation}
  \BDe\BPi(\Bx,\By) = \delta^{(3|3)}(\Bx,\By)
  = \delta^{(3)}(x-y)(\theta-\zeta)^3.
\end{equation}

In the flat upper half space, we can write down simple explicit expressions 
for these propagators. The Laplacian is given by $\BDe=\Delta=\del^\mu\del_\mu$. 
Furthermore we need boundary conditions. We denote by 
$(\tilde\Bx^\mu)=(\tilde x^\mu|\tilde\theta^\mu)$ 
the reflected supercoordinates. Then the boundary for the supercoordinates 
is at $\tilde\Bx=\Bx$. 
Depending on the boundary condition, we find the explicit solution 
\begin{equation}
  \BPi(\Bx,\By) = 
  -\frac{1}{4\pi}\biggl(\frac{(\theta-\zeta)^3}{\|x-y\|} 
 \pm \frac{(\theta-\tilde\zeta)^3}{\|x-\tilde y\|}\biggr), 
\end{equation}
with $+$ ($-$) for Dirichlet (Neumann) boundary conditions. 
Furthermore, we have the explicit form 
$\Bd^\dagger_\Bx=\frac{\del^2}{\del\theta^\mu\del x_\mu}$. 
In terms of these operations we can write the superpropagator as 
\begin{equation}
  \vev{\BX(\Bx)\BF(\By)} 
  = \vev{\Bch(\Bx)\Bps(\By)} 
  =  i\hbar\Bd^\dagger_\Bx\BPi^D(\Bx,\By) =  i\hbar\Bd^\dagger_\By\BPi^N(\Bx,\By). 
\end{equation}

\section{Interactions and Brackets}
\label{sec:interaction}

In this section we will calculate the boundary correlators with a 
single bulk insertion. Note that bulk deformations of order $n$ 
in the conjugate fields (i.e.\ $\psi$ and $F$) give rise to 
$n$-linear brackets on the boundary.

\subsection{A Basic Interaction}

We start with a simple bulk interaction quadratic in the (conjugate) fields, 
\begin{equation}
  \int_\CV \Bps\BF = \int_V (BP+\eta*d\ol\chi). 
\end{equation}
Indeed this is the interaction that should already be turned on 
in the undeformed exact Courant algebroid (and is responsible for 
the Schouten-Nijenhuis bracket on multivector fields). 

This will have an effect on the $AX$ correlator on the boundary. 
It can be motivated formally by noting that $\Bps$ and $\BF$ are the 
conjugate fields to $\Bch$ and $\BX$ respectively. So there is a Feynman 
diagram with the above interaction in the bulk and $\BX$ and 
$\Bch$ on the boundary. 

In fact, as the term above is quadratic it gives a correction to the 
propagators. Let us first, formally, discuss this correction. 
The correction to the 2-point function of $A$ and $X$ is 
\begin{equation}
  \vev{AX}'\sim\frac{i}{\hbar}\vev{AB}\vev{PX} \sim -i\hbar *d_1\Delta_{1,D}\inv d_0\Delta_{0,N}\inv \equiv i\hbar\Xi.
\end{equation}
Naively, this vanishes as we can pull $d_0$ through $\Delta_{1,D}\inv$ 
where it is annihilated by $d_1$. There is however a catch in this argument, 
as the two propagators have different boundary conditions. This makes that 
pulling the $d_0$ through is not allowed. That indeed $\Xi$ is nonzero 
can be seen by acting with $*d$, 
\begin{equation}\label{xieqn}
  *d_1\Xi = (\Delta_1-d_0*d_2*)\Delta_{1,D}\inv d_0\Delta_{0,N}
  = d_0\Delta_{0,N}\inv - d_0\Delta_{0,D}\inv*d_2*d_0\Delta_{0,N}\inv 
  = d_0(\Delta_{0,N}\inv-\Delta_{0,D}\inv). 
\end{equation}
One can show that pulling through $*d_2*$ is allowed because 
the expression is sandwiched in $d_0$'s. Indeed we see that 
the above is nonzero due to the difference in boundary condition. 

Another way to see this equation is to write down the full 
quadratic action for the two coupled vector multiplets when including 
the $\psi F$ term. In matrix notation the relevant part of 
the Lagrangian density can be written as 
\begin{equation}
  \pmatrix{B\cr\ul\chi\cr\ul\eta\cr X}^t *
  \pmatrix{*d_1 & d_0 & 1 & \cdot \cr -*d_2* & \cdot & \cdot & \cdot \cr \cdot & \cdot & *d_1 & d_0 \cr \cdot & \cdot & -*d_2* & \cdot }
  \pmatrix{A\cr\ul\psi\cr *P\cr\ul\lambda}.
\end{equation}
The deformed propagator is the inverse of the kinetic matrix 
appearing above, 
\begin{equation}
  \pmatrix{%
  -*d_1\Delta\inv_{1,D} & -d_0\Delta\inv_{0,N} & -*d_1*d_1\Delta_{1,D}^{-2} & \Xi \cr
  *d_2*\Delta\inv_{1,D} & \cdot & \cdot & \Delta\inv_{0,D} \cr
  \cdot & \cdot & -*d_1\Delta\inv_{1,D} & -d_0\Delta\inv_{0,N} \cr
  \cdot & \cdot & *d_2*\Delta\inv_{1,D} & \cdot},
\end{equation}
with the operator $\Xi$ appearing in the top right corner. The relation 
\eqref{xieqn} is required to get the zero in the top right corner 
of the product of the kinetic matrix with the propagator matrix. 
The other equation $\Xi$ has to satisfy, needed to get a zero 
in the second row, is $d_2*\Xi=0$. This is indeed trivially 
satisfied. We note that in the explicit coordinates the combination 
of scalar propagators appearing in the equation \eqref{xieqn} for $\Xi$,  
$\Pi^{0,N}(x,y)-\Pi^{0,D}(x,y)=\frac{-1}{2\pi}\frac{1}{\|x-\tilde y\|}$, 
only depends on $x-\tilde y$. As a result also the kernel for 
$\Xi$ should depend only on this combination of its arguments. 
This observation will be useful in the explicit calculation of this kernel. 

Let us now be more precise and do the actual calculation of $\Xi$, 
using the explicit form for the propagators given before. 
As these propagators have singularities at coincident points we will need 
some kind of regularization. We will use a point-splitting regularization. 
As the above subtleties suggest the regularization might be important for the result. 
We will comment on this below. The above 2-point function becomes 
\begin{eqnarray}
\vev{A(y)X(x)}' &=& \frac{i}{\hbar}\dint[_V]{d^3\!z} \vev{A(y)B(z)}\vev{P(z)X(x)} \nonumber\\
 &=& -i\hbar\dint[_V]{d^3\!z}d_1\Pi^{1,N}(z,y)d_0\Pi^{0,N}(z,x) \\
 &=& -i\hbar\dint[_{\del V}]{d^2\!z}d_1\Pi^{1,N}(z,y)\Pi^{0,N}(z,x).\nonumber
\end{eqnarray}
The fact that naively this vanishes is reflected by the fact that 
the integrand is a total derivative. However as $V$ has a boundary, 
there remains a boundary term. 
Inserting the propagators we found above, we obtain 
\begin{equation}\label{defprop}
\vev{A_\mu(y)X(x)}' = \frac{-i\hbar}{4\pi^2}\dint[_{\del V}]{d^2\!z} 
  \frac{\eps^{(2)}_{\mu\beta}(y-z)^\beta}{\|y-z\|^3}\frac{1}{\|z-x\|}
  = \frac{i\hbar}{2\pi}\frac{\eps^{(2)}_{\mu\beta}(y-x)^\beta}{\|(x-y)_\parallel\|^2}
  \biggl(1-\frac{x_\perp+y_\perp}{\|x-\tilde y\|}\biggr).
\end{equation}
The explicit calculation of the integral can be found in \appref{app:calc}. 
This result is nonsingular when either $x$ or $y$ are in the bulk. When both 
are on the boundary, it reduces to a simple $1/r$ behavior. This could of 
course have been inferred from the scaling behavior. 

\begin{figure}
\begin{center}
\setlength{\unitlength}{1bp}
\begin{picture}(200,110)
\put(0,0){
}
\thinlines
\put(170,16){\vector(0,1){9}}
\put(170,9){\vector(0,-1){9}}
\put(168,10){\mbox{$\epsilon$}}
\thicklines
\put(20,7){\mbox{$\del V$}}
\put(170,75){\mbox{$z$}}
\put(20,55){\mbox{$V$}}
\put(90,55){\mbox{$\displaystyle\dint{d^3\!z}$}}
\put(100,10){\circle*4}
\put(103,8){\mbox{$x$}}
\put(122,5){\mbox{$\doint{dy}$}}
\end{picture}
\end{center}
\caption{The point splitting regularization, where $x$ and the contour over $y$ are 
taken on the boundary, and the bulk integral is performed over a region 
a distance $\epsilon$ away from the boundary (shaded region).}
\label{region}
\end{figure}
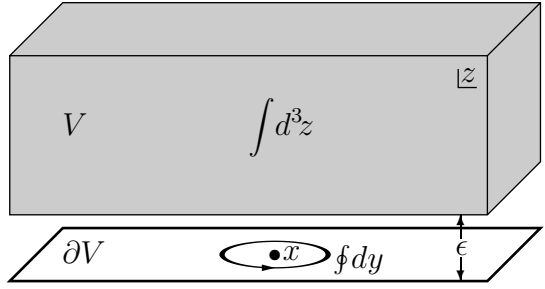

Notice that in the integral above a nonzero $x_\perp$ and $y_\perp$ 
regularize the integral. 
This can be related to the point-splitting regularization. We will be mainly 
interested in the deformed boundary correlators, so we will now take 
$x$ and $y$ on the boundary. The bulk integral over $z$ has to be regularized, 
cutting out small balls around $x$ and $y$. We can alternatively move the 
bulk integration slightly away form the boundary, taking it over $z_\perp>\epsilon$ 
for some $\eps>0$. This is depicted in \figref{region}. We can then safely use 
Stokes's theorem and reduce to the integral over the boundary. Instead of taking 
$z$ off the boundary we can equivalently take $y_\perp$ and $x_\perp$ different from 
zero of order $\epsilon$. We can safely take $x_\perp=0$ as the singularity at 
$z=x$ is harmless. We then end up with the above integral, with $y_\perp$ 
of order $\epsilon$. 

The $1/r$ behavior of the boundary 2-point function implies that in the presence of a 
bulk deformation $\gamma=a_a^i\psi^aF_i$ there is a nonvanishing boundary correlator 
\begin{equation}
 \{\chi_a,X^i\}'\equiv\frac{1}{i\hbar}\Vev{\CO^{(1)}_{\chi_a,C}\CO_{X^i,x}}' = a^i_a, 
\end{equation}
giving the nontrivial bracket. As expected this correlator is independent of the contour $C$ 
(as long as it encloses the point $x$). 
The same calculation is valid for $\vev{*d\ol\eta(y)\chi(x)}$, which gives the other term 
in the bracket, i.e.\ $\frac{1}{i\hbar}\vev{\CO^{(1)}_{X^i,C}\CO_{\chi_a,x}}'=a^i_a$ is also nonzero. 
In the case of the undeformed exact Courant algebroid $A=T^*M$ we had the coupling 
$\gamma=\psi^iF_i$. To find the bilinear bracket on general functions $f,g\in\CB$, 
we substitute observables $\CO^{(1)}_{f,C}$ and $\CO_{g,x}$. As the $\CO^{(1)}_{f,C}$ 
contains just a single field $A$ or $\ol\eta$, there will only be a single deformed 
contraction (factorizing in two undeformed contractions as above). 
With no other contractions, the remaining fields $X$ and $\chi$ 
in the observables will only contribute through their zero modes. 
Therefore, the calculation of the bracket reduces to the calculation of the 
simpler correlator above, see also the discussion around \eqref{bndcorr}. 
The bracket \eqref{bracket} therefore is given by 
\begin{equation}
  \{f,g\}' = (-1)^{|f|+1}\frac{\del f}{\del\chi_i}\frac{\del g}{\del X^i}
  -\frac{\del f}{\del X^i}\frac{\del g}{\del\chi_i},
\end{equation}
where the signs come from the explicit sign in the definition of the bracket 
and commuting $\ol\eta$ through $\frac{\del f}{\del X}$. 
This is precisely the Schouten-Nijenhuis bracket on the boundary algebra 
of multivector fields, $\CB=\Gamma(\ext[]TM)$. 

Let us make a last remark about the above computation. At first sight the result seems 
to be non topological. If we would calculate the correlator above, but with the 
cycle $C$ in the bulk rather than on boundary, we would get a nonzero result 
even though $C$ is now contractable. This correlator however is not topological, 
as the operator $\oint_C A$ is not BRST invariant when $C$ lies inside the bulk, 
as $\BQ\oint_C A=\oint_C d\psi+\oint_C\eta=\oint_C\eta$. Note that on the boundary we do have a 
BRST invariant operator due to the boundary condition on $\eta$. More generally 
as $\BQ P=d\eta$, we could make the observable closed by adding a term 
$-\int_D P$, where $D$ is a disc with boundary $C$. The contribution of this 
extra term will however cancel completely the contribution of the original 
term, making it trivially invariant. We could have added the same contribution 
to the boundary observable. Now the regularization becomes relevant. 
If we regularize by moving the disc slightly in the bulk, the above 
correlator vanishes. However with a point splitting regularization 
we should cut a small hole in the disc. Furthermore, as $P$ vanishes 
on the boundary, the extra contribution is zero and we find the same result as above.

\subsection{Interactions Quadratic in Conjugate Fields}

We will now generalize the above calculation to other interactions still 
quadratic in the conjugate superfields $\BF$ and $\Bps$, but which might include 
extra $\BX$ and $\Bch$ fields. As these are quadratic in conjugate superfields, 
they give contributions to the boundary bracket. We will only 
do the calculation for this bracket. So we consider an interaction of the 
form $\gamma=\varphi_{ab}\psi^a\psi^b$ term in the action, where the 
coefficients $\varphi$ are functions of the 
fields $X$ and $\chi$. To see the effect on the bracket, we insert two 
boundary operators (apart from the outgoing delta-function). 
We will show that 
\begin{equation}\label{quadratic}
  \frac{i}{\hbar}\Vev{\int_\CV(\Bph\Bps\Bps)(\Bz)\, \CO^{(1)}_{\chi,C}\, \CO_{\chi,x}} 
  = i\hbar\varphi(x). 
\end{equation}
Here $C$ is a 1-cycle on the boundary enclosing the point $x$. 
We do not include the indices, as these are obvious. The only important thing will be 
the propagators and the integrals. Explicitly the only contributing term is 
proportional to 
\begin{equation}
  \Vev{\dint[_V]{(\varphi B*d\ol\chi)(z)}\doint[_C]{A(y)}\chi(x)}
  = \dint[_V]{d^3\!z}\doint[_C]{dy}\varphi(z)\Vev{B(z)A(y)}\Vev{*d\ol\chi(z)\chi(x)}, 
\end{equation}
where of course $x$ is on the boundary and $C$ is a cycle on the boundary enclosing $x$. 
At the right hand side we have worked out the contractions that occur. 
Because the $*d\ol\chi\chi$ propagator is the same as the $PX$ propagator, this is actually 
almost the same as the calculation of the bracket we did above. The only difference is 
the presence of $\varphi$. As we saw that the calculation basically reduced 
to local interactions, we should expect that in fact the $z$-dependence of this 
term does not matter, and therefore it can be replaced by $\varphi(x)$. 
This gives exactly the result stated above. 
Let us now confirm that this expectation is correct. 

The presence of the extra factor of $\varphi(z)$ first of all gives an extra factor 
from the partial integration, when the derivative acts on $\varphi$. Furthermore, it 
gives the extra insertion of $\varphi(z)$ in the boundary term. We find, leaving 
out a factor of $\frac{i\hbar}{4\pi^2}$, 
\begin{eqnarray*}
&& \dint[_{\del V}]{d^2\!z}\doint[_C]{dy^\alpha} 
 \varphi(z)\frac{\eps_{\alpha\beta}(y-z)^\beta}{\|z-y\|^3\|z-x\|}  
-\dint[_V]{d^3\!z}\doint[_C]{dy^\alpha} \eps_{\alpha\beta}
 \frac{z^n\del_\beta \varphi(z)+(z-y)^\beta\del_n \varphi(z)}{\|z-y\|^3\|z-x\|} 
\end{eqnarray*}
In the following we will take $x=0$ for simplicity. 
Then we will rescale $z'=z/\|y\|$. 

We start with the second term, which is written 
\begin{equation}
  \dint[_V]{d^3\!z'}\doint[_C]{dy^\alpha} \eps_{\alpha\beta}\biggl(
    \frac{{z'}^n\del_\beta \varphi(\|y\|z')}{\|z'-y'\|^3\|z'\|}
    + \frac{(\|y\|z'-y)^\beta\del_n \varphi(\|y\|z')}{\|y\|\,\|z'-y'\|^3\|z'\|}\biggr).
\end{equation}
Both terms have no pole in $y$, and therefore the contour integral gives zero. 

For the first term we use a Taylor expansion of $\varphi(\|y\|z')$, which becomes 
an expansion in powers of $\|y\|$. We find 
\begin{equation}
  \dint[_{\del V}]{d^2\!z'}\doint[_C]{dy^\alpha} 
   \frac{\eps_{\alpha\beta}(y-\|y\|z')^\beta}{\|y\|^2\|z'-y'\|^3\|z'\|} 
\Bigl(\varphi(0)+\|y\|(z')^\gamma\del_\gamma \varphi(0)+\CO(\|y\|^2)\Bigr). 
\end{equation}
The term of order $\|y\|^2$ has no pole in $y$, and therefore 
the contour integral vanishes. The second term in the expansion gives zero because 
of antisymmetry in the integral under simultaneous reflection of $z$ and $y$. 
The first term in the expansion is the term we are interested in. It is proportional 
to the original integral, and therefore gives $i\hbar\varphi(0)$. 

Substituting back $x$, we conclude therefore that the complete integral equals 
$i\hbar\varphi(x)$, as expected. This calculation shows that a bulk term of the form 
$\frac{1}{2}\int_\CV\Bph_{ab}\Bps^a\Bps^b$ 
---with the coefficients $\varphi_{ab}$ functions of the fields $X$ and $\chi$---induces 
in the deformed boundary theory a contribution to the bracket of the form 
\begin{equation}
  \{f,g\}' = \varphi_{ab}\frac{\del f}{\del\chi_a}\frac{\del g}{\del\chi_b} +\cdots,
\end{equation}
where the ellipses denote contribution from other terms

The calculation above can also be done directly in the superfield notation. 
This has the advantage that several calculations, for different degree forms, 
are done at once.

There is a generalization of the above, which in superfield notation can be written 
\begin{equation}
  \frac{i}{\hbar}\Vev{\int_\CV\Bph\Bps\Bps(\Bz)~ \oint_\CC\Bch(\By)~ \Bch(\Bx)} = i\hbar\Bph(\Bx). 
\end{equation}
The difference with the above is that we did not restrict the dependence of 
the last insertion on $\theta$ (to the zeroth descendant). What we see from this 
is that the combination of the boundary operators behaves like a delta-function 
on the boundary. 

There is a quick way to see this $\delta$-function behavior of the integral. 
Let us shift the integration variables $\Bz$ and $\By$ by $\Bx$ and scale by $R$, 
i.e.\ $\Bz\to R(\Bz-\Bx)+\Bx$. The scaling of the propagators will compensate 
for the scaling of the density. Furthermore, the integral is independent of 
the size of the contour $C$. Therefore the only change is to replace $\Bph(\Bz)$ 
by $\Bph( R(\Bz-\Bx)+\Bx)$. Taking $R\to0$ we find that the full $\Bph$ dependence 
is replaced by $\Bph(\Bx)$. For this argument to work we have to be careful 
that the limit $R\to0$ is continuous. Otherwise, there might be extra terms 
involving derivatives of $\Bph$. Luckily, these terms turn out to vanish. 
Above we showed this was correct for derivatives with respect to $z$. 
For derivatives with respect to $\xi$ a similar calculation will give 
the more general result. In this paper we will actually not need this 
more general result, so we do not give the full derivation.

\subsection{Interactions Cubic in Conjugate Fields}

Next we consider the interactions that are cubic in conjugate fields, 
i.e. of the form $\gamma=\frac{1}{3!}c_{abc}(X,\chi)\psi^a\psi^b\psi^c$. 
Having three conjugate fields $\psi$ we need to insert three 
boundary observables involving $\chi$. This leads to a correlator of the form 
\begin{equation}\label{cubic}
  \frac{i}{6\hbar}\Vev{\int_\CV(\Bc\Bps\Bps\Bps)(\Bz)~\CO^{(1)}_{\chi,C}~\CO_{\chi,x}~\CO^{(2)}_{\chi,\del V}}.
\end{equation}
In fact, we can use the result above for the quadratic interactions 
to calculate this seemingly more complex correlator. 
After the contractions we can write this correlator as 
\begin{equation}
  \frac{i}{\hbar}\dint[_V]{d^3\!z}\dint[_{\del V}]{d^2\!u}\doint[_C]{dy}c(z)
  \vev{\psi(z)*d\ol\psi(u)}\,\vev{B(z)A(y)}\,\vev{*d\ol\chi(z)\chi(x)}. 
\end{equation}
This has indeed the form of the correlator in \eqref{quadratic} with 
$\Bph(\Bz)$ replaced by 
\begin{equation}
  \Bph(\Bz) = \frac{i}{\hbar}\dint[_{\del V}]{d^2\!u}c(z)\vev{\psi(z)*d\ol\psi(u)}. 
\end{equation}
The result \eqref{quadratic} then gives for the above correlator 
\begin{equation}
  i\hbar\varphi(x) 
  = -\frac{\hbar^2c(x)}{2\pi} \dint[_{\del V}]{d^2u}\frac{u_\perp}{\|u-x\|^3}
  = -\hbar^2c(x)\dint[_0^\infty]{dr}\frac{r}{(1+r^2)^{3/2}} = -\hbar^2c(x), 
\end{equation}
where $r=\frac{\|u_\parallel-x_\parallel\|}{|u_\perp|}$. 
Note that here we need $u_\perp>0$, which is satisfied because of the 
point-splitting regularization. It follows that the correlator 
\eqref{cubic} is equal to $-\hbar^2 c(x)$. 
Here the coefficient $c$ can again be any function of the fields $X$ and $\chi$. 
The correlator \eqref{cubic} calculates the trilinear bracket in \eqref{brackets}. 
As for the bilinear bracket, for general arguments $f,g,h\in\CB$ of this bracket, 
the calculation reduces essentially to the above correlator, with some 
extra signs coming the definition \eqref{brackets} and from 
straightforward ordering of the factors. We conclude that the interaction 
term $\frac{1}{3!}\int_\CV\Bc_{abc}\Bps^a\Bps^b\Bps^c$ gives a 
contribution to the trilinear bracket of the form 
\begin{equation}
  \{f,g,h\}' = c_{abc}\frac{\del f}{\del\chi_a}\frac{\del g}{\del\chi_b}\frac{\del h}{\del\chi_c} +\cdots,
\end{equation}
with the ellipses again denoting contributions from other terms.

\subsection{Boundary Closed String Field Theory}

The $L_\infty$ brackets we calculated through the correlation functions 
generate the closed string field theory action of the boundary string. 
Indeed, the $L_\infty$ algebra of the bosonic closed string field theory  
of \cite{zwie} will be the same as the $L_\infty$ algebra discussed in 
this paper for the present topological situation. 
The structure constants of this $L_\infty$ algebra, together with the 
natural pairing defined by the 2-point functions, can be interpreted 
as the coefficients of an action functional for the closed string 
field theory \cite{zwie}. Therefore, we have 
basically calculated the  string field theory action to lowest order 
for the boundary string theory of the open membrane. 

For the models discussed in this paper, the string field 
of the boundary closed string field theory is an element 
$\Phi$ living in the boundary algebra $\CB=C^\infty(\CL)$. 
The inner product is defined in terms of the 2-point function. 
This can be reduced to an integral over the zero modes, 
which are the coordinates $(X^i,\chi_a)$ on the supermanifold $\CL$. 
The string field theory action is given in terms of the brackets by 
\begin{equation}
  S = \int_\CL \Bigl( \frac{1}{2} \Phi Q\Phi + \frac{1}{3}\Phi\{\Phi,\Phi\} + \frac{1}{4}\Phi\{\Phi,\Phi,\Phi\} +\cdots \Bigr).
\end{equation}

Let us summarize the results for the case of the 3-form model, 
based on the target superspace $\CM=T^*[2](\Pi T^*M)$. In this 
case the boundary algebra $\CB=C^\infty(\Pi T^*M)=\Gamma(M,\ext TM)$ 
can be identified with the algebra of polyvector fields. 
The basic string field is related to a bivector 
$\Phi=\frac{1}{2}B^{ij}(X)\chi_i\chi_j$, the other components 
correspond to ghosts and antifields in the closed string field theory. 
The deformations \eqref{cgamma} were based on a closed 3-form 
$c_{ijk}$ and a quasi-Poisson bivector $b^{ij}$. 
We conclude that the induced $L_\infty$ structure is given to first 
order in $c$ by 
\begin{eqnarray}
  Q &=& b^{ij}\chi_j\frac{\del}{\del X^i}+\frac{1}{2}(\del_kb^{ij}+c_{klm}b^{li}b^{mj})\chi_i\chi_j\frac{\del}{\del\chi_k} +\CO(c^2),\nonumber\\
  \{\cdot,\cdot\} &=& \frac{\del}{\del X^i}\wedge\frac{\del}{\del\chi_i} +\frac{1}{2}c_{ijk}b^{kl}\chi_l\frac{\del}{\del\chi_i}\wedge\frac{\del}{\del\chi_j} +\CO(c^2),\\
  \{\cdot,\cdot,\cdot\} &=& \frac{1}{6}c_{ijk}\frac{\del}{\del\chi_i}\wedge\frac{\del}{\del\chi_j}\wedge\frac{\del}{\del\chi_k} +\CO(c^2).\nonumber
\end{eqnarray}
We recognize in the undeformed bracket the Schouten-Nijenhuis bracket 
on polyvector fields. Furthermore, for $c=0$ the BRST operator $Q$ 
is the standard Poisson differential for the Poisson structure $b^{ij}$. 
We also note that the essential deformation due to the 3-form 
$c$ is the trilinear bracket, corresponding to a cubic interaction 
in the string field theory action.

\section{Conclusions and Discussion}
\label{sec:concl}

We calculated the propagators and correlators for the topological 
open membrane to leading order. This leads to a confirmation 
that the $L_\infty$ algebra of the boundary string theory indeed 
is given by the algebraic expressions related to the bulk 
couplings parameters. 

We note that adopting the point-splitting regularization 
is essential in getting precisely this result. It is 
similar to what happens in the case of strings coupling to a 2-form 
\cite{seiwit}. For the string theory, the bulk interaction 
induces a 2-point function with a step-function behavior. This 
results in the noncommutativity of the product in the boundary 
open string theory \cite{seiwit}. In our case we find a 
deformed 2-point function that has a $1/r$ behavior on the 
boundary. This shows that rather the bracket is deformed. 

In fact as the deformation has degree 3, the generic deformation 
will generically contain a term of order 3 in the conjugate fields. 
As our calculations show, this gives rise to the trilinear bracket. 
In the quantized algebra, this integrates to a Drinfeld associator. 
This is most clearly seen in the simplest situation where we have 
only degree 1 multiplets $(\chi_a,\psi^a)$. The semiclassical 
trilinear bracket we have calculated is induced from the coefficients 
$c_{abc}$ of the deformation. As shown in \cite{tomalg} this 
situation corresponds to a quasi-Lie bialgebra, with these 
coefficients identified with the structure constants of the coassociator. 
This indeed is the infinitesimal structure of the Drinfeld associator 
\cite{drin1} of a quasi-Hopf algebra. 

The other special situation is the exact Courant algebroid, 
for which $\CM=T^*[2](T^*M)$. It is the toy model of a string 
in a 3-form background. We found that it gives rise to a closed 
string field theory with a quartic coupling proportional to 
the 3-form. This coupling is equivalent to the trilinear bracket 
in the $L_\infty$ algebra. The open string version gives rise to 
a deformation of the problem of deformation quantization, 
as discussed in \cite{tomalg,sevwein}. 

The quantization performed in this paper is only the first step 
towards a quantization of the open membrane. Indeed, the calculations 
we did here only reproduced the semi-classical quasi-Lie bialgebroid 
structure. The full quantum correlators, involving higher orders in 
the bulk coupling, will give a quantization of this quasi-Lie 
bialgebroid structure, what could be called a quasi-Hopf algebroid. 
The $L_\infty$ structure we found will integrate the product structure 
in this object. This idea is more easily tested in the much better 
understood case of quasi-Lie bialgebras, when only the degree 1 
multiplets are present. The full quantum result in this case should 
reflect the quasi-Hopf algebra structure. As these models are 
of a Chern-Simons type, this relation can be viewed as a generalized 
Chern-Simons/WZW correspondence \cite{witpol}. We will leave 
the study for further quantization of these open membrane models 
and the emergence of the quasi-Hopf algebra structure for a 
later paper \cite{tomquant}.

\section*{Acknowledgements}

We are happy to thank Hong Liu, Jeremy Michelson, Sangmin Lee, David Berman 
and Jan-Pieter van der Schaar for interesting discussions. 
The research of C.H. was partly supported by DOE grant \#DE-FG02-96ER40959; 
J.-S.P. was supported in part by NSF grant \#PHY-0098527 and by the Korea Research Foundation.

\appendix

\section{Calculation of the Integral}
\label{app:calc}

In this appendix we calculate explicitly the integral in \eqref{defprop}. 
We shift $z$ along $x_\parallel$ (parallel to the boundary), 
rescale $z$ by $\|(x-y)_\parallel\|$ 
and change to polar coordinates. We see that the only 
surviving component is the one perpendicular 
to the direction of $(x-y)_\parallel$, which is given by 
\begin{equation}
\frac{-i\hbar}{4\pi^2\|(x-y)_\parallel\|}\dint[_0^\infty]{dr}\dint[_0^{2\pi}]{d\phi}
  \frac{r^2\cos\phi-r}{(r^2-2r\cos\phi+1+\eta^2)^{3/2}(r^2+\xi^2)^{1/2}},
\end{equation}
where $\eta = \frac{y_\perp}{\|(x-y)_\parallel\|}$ and $\xi =\frac{x_\perp}{\|(x-y)_\parallel\|}$. 
In general this integral can not be calculated exactly. 
However, when we take $x$ on the boundary, we can explicitly perform the integral. 
The integral, without the factor $\frac{-i\hbar}{4\pi^2\|(x-y)_\parallel\|}$ 
in front, reduces to 
\begin{eqnarray}
 \dint[_0^{2\pi}]{d\phi}\dint[_0^\infty]{dr}\frac{r\cos\phi-1}{(r^2-2r\cos\phi+1+\eta^2)^{3/2}} 
  &=& \dint[_0^{2\pi}]{d\phi}\frac{-r\sin^2\phi-\eta^2\cos\phi}{(\sin^2\phi+\eta^2)\sqrt{r^2-2r\cos\phi+1+\eta^2}}\Biggr|_{r=0}^\infty \nonumber\\
  &=& \dint[_0^{2\pi}]{d\phi}\frac{-1}{\sin^2\phi+\eta^2}\biggl(\sin^2\phi-\frac{\eta^2\cos\phi}{\sqrt{\eta^2+1}}\biggr) \\
  &=& -2\pi\biggl(1-\frac{\eta}{\sqrt{\eta^2+1}}\biggr)
  = -2\pi\biggl(1-\frac{y_\perp}{\|x-y\|}\biggr).\nonumber
\end{eqnarray}
We now use the fact that the correlator only depends on the 
combination $x-\tilde y$, or equivalently $y-\tilde x$. 
This also means that it depends on the normal 
coordinates only through $x_\perp+y_\perp$. This allows us to find the 
full answer for nonzero $x_\perp$, simply by replacing $y-x$ by $y-\tilde x$ 
and $y_\perp$ by $x_\perp+y_\perp$.

\end{document}